\documentclass[11pt, letter]{article}
\usepackage{jheppub}
\usepackage[english]{babel}
\usepackage{blindtext}\blindmathtrue
\usepackage{avant} % Use the Avantgarde font for headings
\usepackage[dvipsnames]{xcolor}
\usepackage{amsmath,epsf,amssymb,mathtools,latexsym,amsthm,setspace,array,pifont,hyperref,amsfonts,dsfont,cancel,braket,parskip}
\usepackage[none]{hyphenat} % No hyphenated line endings
\usepackage{graphicx}
\usepackage{float}
\usepackage{tikz} 
\usepackage{tikz-feynman, contour}
\tikzfeynmanset{compat=1.1.0}
\usetikzlibrary{shapes,arrows,positioning,automata,backgrounds,calc,er,patterns}
\makeatletter
\newlength\CoolS@sizex
\newlength\CoolS@sizey
\newcommand*\CoolS@inner{%
\begin{tikzpicture}[baseline=0.04\CoolS@sizey]%
\foreach \x in {0, 1, ..., 5} \foreach \y in {0, 1, ..., 10}
\coordinate (c\x\y) at (\x *0.12*\CoolS@sizex, \y *0.107*\CoolS@sizey);
\draw [line width=\Cool@stroke] (c28)--(c26)--(c44)--(c42)--(c20)--(c02)--(c04)--(c15);
\draw [line width=\Cool@stroke] (c22)--(c24)--(c06)--(c08)--(c210)--(c48)--(c46)--(c35);
\end{tikzpicture}}
\usepackage{todonotes}
\graphicspath{{Pictures/}} % Specifies the directory where pictures are stored
\newcommand{\nn}{\nonumber}
\newcommand{\sd}{\mathrm{d}}

\newcommand{\cl}[1]{\mathcal{#1}}

\renewcommand{\to}{\longrightarrow}

\def\prd{\ref@{Phys.~Rev.~D}}        % Physical Review D

\newcommand{\td}[1]{
\if\notesOn1
\todo[inline]{#1}
\fi
}
\hypersetup{
	colorlinks = true,
	linkcolor = Mahogany,
	anchorcolor = Mahogany,
	citecolor = Mahogany,
	filecolor = Mahogany,
	urlcolor = Mahogany
}
\newcommand{\Li}{\text{Li}}
\def\notesOn{1}
\tikzset{
	graviton/.style={
		double,
		decoration={snake, aspect=0.75, mirror, segment length=1.5mm},
		decorate
	}
}

\tikzfeynmanset{
	bigblob/.style={
		shape=circle,
		draw=blue,
		fill=red}
}

\title{Scattering Amplitudes, Black Holes and Leading Singularities in Cubic Theories of Gravity}
\author[\lambda] {William T. Emond}
\author[\alpha]{ and Nathan Moynihan}

\affiliation[\lambda]{School of Physics and Astronomy, University of Nottingham,
University Park, Nottingham NG7 2RD, United Kingdom}
\affiliation[\alpha]{The Laboratory for Quantum Gravity \& Strings, Department of Mathematics \& Applied Mathematics, University Of Cape Town}
\emailAdd{william.emond@nottingham.ac.uk}
\emailAdd{nathantmoynihan@gmail.com}
\abstract{
	We compute the semi-classical potential arising from a generic theory of cubic gravity, a higher derivative theory of spin-2 particles, in the framework of modern amplitude techniques. We show that there are several interesting aspects to this potential, including some non-dispersive terms that lead to black hole solutions (including quantum corrections) that agree with those derived in Einsteinian cubic gravity (ECG). We show that these non-dispersive terms could be obtained from theories that include the Gauss-Bonnet cubic invariant $G_3$. In addition, we derive the one-loop scattering amplitudes using both unitarity cuts and via the leading singularity, showing that the classical effects of higher derivative gravity can be easily obtained directly from the leading singularity with much less computational cost.
}

\begin{document}

\maketitle
\section{Introduction}
The modern S-matrix program has been wildly successful when applied to gravitation \cite{Bedford:2005yy,Nguyen:2009jk,Benincasa:2007qj,Bjerrum-Bohr:2018xdl}, including higher derivative theories \cite{Johansson:2017xj,He:2016iqi,Dunbar:2017dgp,Dunbar:2017szm,Carballo-Rubio:2018bmu}. In the age of LIGO, there has been much attention in the simplicity of amplitude techniques when computing post-Newtonian and post-\; Minkowskian corrections to General Relativity (GR) \cite{Cheung:2018wkq,Bern:2019nnu}, and it has been shown that the classical contribution of loop amplitudes correspond to terms in a post-Minkowskian expansion \cite{Damour_2016,Damour_2018}. As an interesting modification of GR, one can consider theories of gravity which involve cubic terms in either the Riemann or Ricci tensors which, among other things, contain non-trivial black hole solutions in four dimensions \cite{Bueno:2016lrh}. These higher-derivative contributions to the gravitational action are often encountered within string theory \cite{TSEYTLIN1986391} and can be formulated in this way as they possess only spin-2 degrees of freedom on-shell \cite{Bueno:2016xff, Gullu:2014gza}. In this paper, we will explore both the classical and quantum aspects of this class of theories by computing on-shell scattering amplitudes. From there, we will use these amplitudes to derive the semi-classical potential associated with cubic theories of gravity where the graviton mediated interaction between two scalars is affected by cubic terms only at one-loop order and above. 

In section \ref{cubic} we review cubic theories of gravity -- including Einsteinian cubic gravity -- in order to set up the problem we will consider. In section \ref{potentials}, we develop the tools required to obtain the semi-classical potential and black hole solutions directly from scattering amplitudes before moving on to section \ref{unitaritycuts}, where we compute the massive scalar one-loop amplitude using unitarity cuts. This requires us to compute the coefficients of the standard integrals that usually arise in a Passarino-Veltman loop decomposition. From this, we derive the quantum-\; corrected classical potential and the classical and quantum corrections to the Schwarzschild black hole solution arising from the addition of a cubic term in the gravitational action.  Computing amplitudes via unitarity cuts is computationally expensive and so, to contrast this, in section \ref{leadingS} we derive the classical contribution of the amplitude directly from the \textit{Leading singularity} \cite{Cachazo:2017jef}, where loop integration is reduced to the far simpler problem of computing residues.

\section{Cubic Theories of Gravity}\label{cubic}
Higher derivative operators in gravity are important for a variety of reasons, including the modification of gravity at short distances/large energies and the possibility of renormalisability. One particularly interesting theory of gravity in this class is \textit{Einsteinian cubic gravity} (ECG) \cite{Bueno:2016xff,Bueno:2016lrh}, which enjoys the same linearised spectrum as General Relativity, in that it propagates only two degrees of freedom on-shell.

We will consider a generic six-derivative theory in four dimensions described by the action
\begin{equation}\label{key}
S = \int \sd^4x\sqrt{-g} \left(\frac{2}{\kappa^2}R + \lambda \mathcal{P}\right),
\end{equation}
where the coupling has mass dimension $[\lambda] = -2$ and
\begin{equation}
\mathcal{P}=\beta_1 R^{\mu}_{\ \ \alpha\nu\beta}R^{\alpha \lambda\beta\sigma}R_{\lambda\mu\sigma}^{\ \ \ \ \nu}+\beta_2 R_{\mu\nu}^{\ \ \ \alpha\beta}R_{\alpha\beta}^{\ \ \ \lambda\sigma}R_{\lambda\sigma}^{\ \ \ \mu\nu} + \beta_3 R_{\mu\nu\alpha\beta}R^{\mu\alpha}R^{\nu\beta} + \beta_4 R_{\mu}^{\ \ \nu}R_{\nu}^{\ \ \alpha}R_{\alpha}^{\ \ \mu}.
\end{equation}
We leave these coefficients generic, in order to keep track of how each of these terms contributes to the physical effects, however, when required to specialise to ECG, we will consider the specific set of coefficients
\begin{equation}
    \beta_1 = 12,\; \beta_2 = 1,\; \beta_3 = -12,\; \beta_4 = 8.
\end{equation}
Nonetheless, as is well known, only the first two terms typically contribute to the S-matrix at cubic order and, furthermore, a specific choice of coefficients, $\beta_1 = -2\beta_2$, gives the well known cubic Gauss-Bonnet invariant
\begin{equation}
    G_3 = R_{\mu\nu}^{\ \ \ \alpha\beta}R_{\alpha\beta}^{\ \ \ \lambda\sigma}R_{\lambda\sigma}^{\ \ \ \mu\nu} -2R^{\mu}_{\ \ \alpha\nu\beta}R^{\alpha \lambda\beta\sigma}R_{\lambda\mu\sigma}^{\ \ \ \ \nu}.
\end{equation}
While this term does not produce pure graviton dynamics on its own, when coupled to Einstein gravity or generic matter, it can produce non-trivial scattering effects \cite{1987PhLB..185...52M,Broedel:2012rc}.
As expected from a cubic theory of gravity, and with the predictable coefficients given the argument above, the on-shell three-point all minus graviton amplitude at order $\lambda$ is given by
\begin{equation}\label{key}
M^{---} = \frac{3}{8}\kappa^3\lambda(\beta_1+2\beta_2)\braket{12}^2\braket{23}^2\braket{31}^2,
\end{equation}
where we have derived this using eq.~\eqref{negcurrent} contracted with a graviton polarisation tensor.
At tree level, when compared with the contributions from General Relativity (GR), we find that although the three-point vertex itself is modified by the $\cl{O}(R^3)$ terms, we do not modify the scalar-scalar-graviton vertex. This means that, at first-order in $G$, the Newtonian potential must be the same in both GR and ECG, and thus we would expect any static, spherically symmetric black hole solutions to be a higher-order perturbation of the Schwarzschild solution. To find such a higher order contribution we must compute the classical contributions that arise from a one-loop amplitude and, from this, derive the classical potential.

\section{Scattering Amplitudes and the Effective Potential}\label{potentials} 
A particularly sensible definition of the potential energy is to define it in terms of gauge-invariant on-shell scattering amplitudes in the non-relativistic limit. To this end, we will consider $2\longrightarrow 2$ scattering of two massive scalars mediated by gravity. This ensures that the definition of the potential is itself gauge-invariant \cite{Kazakov_2001} and that, in the non-relativistic limit ($t = -\mathbf{q}^2$), can be given by the inverse Born approximation \cite{Iwasaki_1971,Iwasaki_1971B}
\begin{equation}\label{key}
V(\textbf{r},\textbf{p}) = -\frac{1}{4E_AE_B}\int \frac{\sd^3\mathbf{q}}{(2\pi)^3}e^{i\mathbf{q}\cdot \mathbf{r}}\cl{M}(\textbf{q}, \textbf{p}),
\end{equation}
where $\textbf{q}$ is the exchanged three-momentum and $E_A ~(E_B)$ is the energy associated with particle $A~ (B)$.
\begin{figure}[H]
	\centering
	\begin{equation}\label{setup}
	\begin{tikzpicture}[scale=0.75]
	\begin{feynman}
	\vertex (m) at (0, 0);
	\vertex (mp) at (0,0);
	\vertex (x) at (4,2) {$P_4$};
	\vertex (y) at (4,-2) {$P_3$};
	\vertex (a) at (-4,2) {$P_1$};
	\vertex (b) at (-4,-2) {$P_2$};
	%\vertex (c) at (4,-2) {$P_3$};
	%\vertex (d) at (4,2) {$P_4$};
	\diagram* {
		(a) -- [plain] (m) -- [plain] (x),
		(b) -- [plain] (mp) -- [plain] (y)
	};
	\draw[preaction={fill, white},pattern=north east lines] (0,0) ellipse (1.2cm and 1.2cm);
	\draw (-2,0) node {$m_A$};
	\draw (2,0) node {$m_B$};
	\end{feynman}
	\end{tikzpicture} \nn
	\end{equation}
	\caption{The kinematic setup, where particles 1 and 4 are incoming and 2 and 3 outgoing. In the center of mass frame, we consider the exchanged momentum $q = P_1+P_2 = (0,\textbf{q})$ (in the all outgoing convention) and $P_1 = (E_A,\textbf{q}/2)$, $P_2 = -(E_A,-\textbf{q}/2)$ with $E_A = \sqrt{m_A^2 + \textbf{p}^2 + \frac{\textbf{q}^2}{4}}$. $P_3$ and $P_4$ are defined similarly with $A\leftrightarrow B$.}
\end{figure}
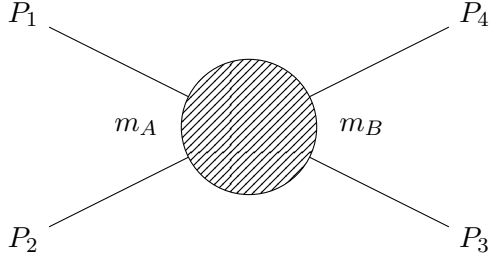

Ultimately, we would like to derive the metric associated to a black hole in an asymptotically flat spacetime \cite{PhysRevD.9.1837,Neill_Rothstein_2013}, meaning we need to relate the potential energy to the metric. We will therefore consider the gravitational field to be sourced by two point-masses in the stationary limit. In this limit, the usual relativistic action for a point particle is only dependent on $g_{00}$ and we therefore consider the following path integral \cite{Modanese_1995}
\begin{equation}\label{key}
Z = \int \cl{D}h_{\mu\nu}~\exp\left[{-i\left(S_{EH} + m_A\int_{-T/2}^{T/2}d\sigma \sqrt{g_{00}(\sigma)} + m_B\int_{-T/2}^{T/2}d\sigma' \sqrt{g_{00}(\sigma')}\right)}\right].
\end{equation}
This describes the two sources, with masses $m_A$ and $m_B$, interacting at rest on a flat background. They begin at some fixed distance, with their interaction adiabatically turned on at a (large) time $-T/2$, and turned off at $T/2$. We can then consider the generating functional given by
\begin{align}\label{key}
\cl{F} &= \frac{\int \cl{D}h_{\mu\nu}~\exp\left[{-i\left(S_{EH} + m_A\int_{-T/2}^{T/2}d\sigma \sqrt{g_{00}(\sigma)} + m_B\int_{-T/2}^{T/2}d\sigma' \sqrt{g_{00}(\sigma')}\right)}\right]}{\int \cl{D}h_{\mu\nu}~\exp\left[{-iS_{EH}}\right]}\nn\\[1em]
&= \left\langle{\exp\left[{-i\left(m_A\int_{-T/2}^{T/2}d\sigma \sqrt{g_{00}(\sigma)} + m_B\int_{-T/2}^{T/2}d\sigma' \sqrt{g_{00}(\sigma')}\right)}\right]}\right\rangle.
\end{align}
In the $T\longrightarrow\infty$ limit, this is well approximated by the ground state energy \cite{Modanese_1995}, meaning we can say that $\lim_{T\longrightarrow\infty}\cl{F} \sim e^{-iV(R)T}$, and we can define the potential energy via
\begin{equation}\label{potint}
V(r) = \lim_{T\longrightarrow\infty}\frac{i}{T}\log(\cl{F}) \simeq \lim_{T\longrightarrow\infty}\frac{1}{T}\left(m_A\int_{-T/2}^{T/2}d\sigma \sqrt{g_{00}(\sigma)}  + m_B\int_{-T/2}^{T/2}d\sigma' \sqrt{g_{00}(\sigma')}\right).
\end{equation}
In the static, spherically symmetric limit in which we are interested, the potential $\Phi$ is related to the metric via
\begin{equation}\label{metricpotential}
g_{00} = 1 - 2\Phi.
\end{equation}
If we consider the probe limit where $m_B\ll m_A$, then we can discard the gravitational field produced by $m_B$ and easily perform the integral in eq.~\eqref{potint} for the static case (where we take $|\Phi| \ll 1$), to find
\begin{equation}\label{key}
V(r) = m_B\sqrt{1-2\Phi} \simeq m_B(1-\Phi).
\end{equation}
Deriving the potential energy $V(r)$ from the amplitudes allows us to compute the potential directly. One way to do this is to expand $\Phi(r,m_A)$ in terms of $G$ as
\begin{equation}\label{key}
\Phi(r,m_A) = \sum_{n=1}^\infty C_n(m_A,r) G^n,
\end{equation}
where $C_n(r,m_A)$ will be a combination of $m_A$ and $r$ with mass dimension $2$. We can then directly compare this order-by-order with the potential energy in the correct limit  
\begin{equation}\label{key}
\Phi(r,m_A) = \sum_{n=1}^\infty C_n(r,m_A) G^n = -\lim_{m_B\longrightarrow 0}\frac{1}{m_B}\left(\frac{1}{m_Am_B}\int \frac{\sd^3\mathbf{q}}{(2\pi)^3}e^{i\mathbf{q}\cdot \mathbf{r}}\cl{M}(\textbf{q})\right).
\end{equation}
With this solution in hand, the line element is given by
\begin{equation}\label{key}
ds^2 = -(1-2\Phi)dt^2 + (1+2\Phi)dr^2 + r^2 d\Omega.
\end{equation}

\section{One Loop Amplitude} 
\subsection{Unitarity Cuts}\label{unitaritycuts}
In order to see any effects from cubic gravity, we are required to compute at minimum a one-loop amplitude. This is due to the fact that we require the presence of the three-point same-helicity amplitude, which cannot occur during a tree-level interaction. We will first compute the relevant one-loop amplitude using standard on-shell unitarity cuts. As is well known, corrections to the potential arise from the purely non-analytic pieces of loop amplitudes corresponding to long-range effects of massless particle interactions \cite{Holstein:2004dn}. This means that we only need to consider cuts in the $t$-channel, and we need not consider all possible cuts. Indeed, we are free to ignore those that will give purely analytic contributions to the amplitudes. The non-analytic pieces of loop amplitudes are also independent of regularisation scheme, and as such we can happily work in $D=4$ throughout the calculation \cite{1994PhRvD..50.3874D}. Since only the graviton three-point vertex is modified in cubic theories, the one-loop box diagram must be the same as it is in GR, and thus we will focus first on the triangle diagram, noting that any contributions ought to come from diagrams containing massive propagators, which facilitate the delicate $\hbar$ cancellations that give rise to purely classical pieces \cite{Holstein:2004dn, Kosower:2018adc}. We will consider the following diagram
\begin{figure}[H]
	\centering
	\begin{tikzpicture}[scale=1]
	\begin{feynman}
	\vertex (m) at ( -2, 0);
	\vertex (mp) at ( -2, -0.5);
	\vertex (q) at ( 2, 0);
	\vertex (qp) at (2,-0.5);
	\vertex (x) at (-0.1,0);
	\vertex (xp) at (0.1,0);
	\vertex (y) at (-0.1,-0.5);
	\vertex (yp) at (0.1,-0.5);  
	\vertex (a) at (-4,2) {$P_1$};
	\vertex (b) at (-4,-2) {$P_2$};
	\vertex (c) at (4,-2) {$P_3$};
	\vertex (d) at (4,2) {$P_4$};
	\diagram* {
		(a) -- [plain] (m) -- [graviton] (x) (xp) -- [graviton] (q) -- [plain] (d),
		(b) -- [plain] (mp) -- [graviton] (y) (yp) -- [graviton] (qp)  -- [plain] (c)
	};
	\draw[preaction={fill, white},pattern=north east lines] (-1.9,-0.25) ellipse (0.3cm and 0.8cm);
	\draw[preaction={fill, white},pattern=north east lines] (1.9,-0.25) ellipse (0.3cm and 0.8cm);
	\draw [densely dashed, red, line width=0.3mm,] (0,1.2) -- (0,-1.6);
	\draw (-1,0.1) node[above] {$\ell_1 \longrightarrow$};

	\draw (-1,-0.65) node[below] {$\ell_2 \longrightarrow$};
	\end{feynman}
	\end{tikzpicture}
		\caption{Double Cut Diagram}
\end{figure}
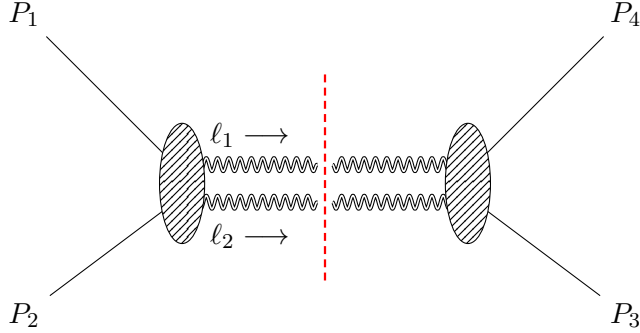
%\bigskip\bigskip
To compute the double cut, we need to evaluate
\begin{equation}\label{key}
M_4^{(1)} = -i\sum_{h_1,h_2}\int \frac{\sd^4\ell_1}{(2\pi)^4}~\frac{M_L[P_1,P_2,\ell_1^{h_1},\ell_2^{h_2}]M_R[-\ell_1^{-h_1},-\ell_2^{-h_2},P_3,P_4]}{\ell_1^2\ell_2^2}\Bigg|_{\ell_1^2 = \ell_2^2 = 0},
\end{equation}
where $\ell_2 = \ell_1 - P_1 - P_2$ and $P_1^2=P_2^2 = m_A^2$, $P_3^2 = P_4^2 = m_B^2$. The cut conditions are therefore given by
\begin{flalign}\label{eq:cut conditions}
&\ell_{1}^{2} \ = \ (\ell_{1} \: - \: q)^{2} \ = \ 0 \nonumber\\[0.8em] &\Rightarrow\qquad 2\ell_{1}\cdot q \ = \ q^{2} \; ,
\end{flalign}
where $q=P_{1}+P_{2}$. Note that eq.~\eqref{eq:cut conditions} implies that $P_{2}\cdot\ell_{1}=m_{A}^{2}+P_{1}\cdot P_{2}-P_{1}\cdot\ell_{1}=\frac{t}{2}-P_{1}\cdot\ell_{1}$. Moreover, we define the Mandelstam variable $t=q^{2}=(P_{1}+P_{2})^{2}=(\ell_{1}+\ell_{2})^{2}=2\ell_{1}\cdot\ell_{2}$ (we adopt the all-outgoing convention for external particle momenta). The tree level diagrams on both sides of the cut are the classical gravitational Compton diagrams, given by 
\begin{figure}[H]
	\centering
	\begin{equation}\label{gravcompton}
	\begin{gathered}
	\begin{tikzpicture}[scale=1]
	\begin{feynman}
	\vertex (m) at ( -2, 0);
	\vertex (mp) at ( -2, -0.5);
	\vertex (q) at ( 2, 0);
	\vertex (qp) at (2,-0.5);
	\vertex (x) at (-0.1,0);
	\vertex (xp) at (0.1,0);
	\vertex (y) at (-0.1,-0.5);
	\vertex (yp) at (0.1,-0.5);  
	\vertex (a) at (-4,2) {$P_1$};
	\vertex (b) at (-4,-2) {$P_2$};
	%\vertex (c) at (4,-2) {$P_3$};
	%\vertex (d) at (4,2) {$P_4$};
	\diagram* {
		(a) -- [plain] (m) -- [graviton] (x),
		(b) -- [plain] (mp) -- [graviton] (y)
	};
	\draw[preaction={fill, white},pattern=north east lines] (-1.9,-0.25) ellipse (0.3cm and 0.8cm);
	%\draw[preaction={fill, white},pattern=north east lines] (1.9,-0.25) ellipse (0.3cm and 0.8cm);
	%\draw [densely dashed, red, line width=0.5mm,] (0,1.2) -- (0,-1.6);
	\draw (-1,0.1) node[above] {$\ell_1 \longrightarrow$};
	\draw (-1,-0.65) node[below] {$\ell_2 \longrightarrow$};
	\end{feynman}
	\end{tikzpicture}
	\end{gathered}
	\begin{gathered}
	~~~=~~~
	\end{gathered}
	\begin{gathered}
	\begin{tikzpicture}[scale=0.6]
	\begin{feynman}
	\vertex (m) at ( -2, 1);
	\vertex (mp) at ( -2, -1);
	\vertex (x) at (-0.1,1) {$\ell_1$};
	\vertex (y) at (-0.1,-1) {$\ell_2$}; 
	\vertex (a) at (-4,2) {$P_1$};
	\vertex (b) at (-4,-2) {$P_2$};
	\diagram* {
		(a) -- [plain] (m) -- [graviton] (x),
		(b) -- [plain] (mp) -- [graviton] (y),
		(m) -- [plain] (mp)
	};
	\end{feynman}
	\end{tikzpicture}
	\end{gathered}
	~~+~
	\begin{gathered}
	\begin{tikzpicture}[scale=0.6]
\begin{feynman}
\vertex (m) at ( -3, 0);
\vertex (mp) at ( -2, 0);
\vertex (x) at (-0.1,1) {$\ell_1$};
\vertex (y) at (-0.1,-1) {$\ell_2$}; 
\vertex (a) at (-4,2) {$P_1$};
\vertex (b) at (-4,-2) {$P_2$};
\diagram* {
	(a) -- [plain] (m) -- [graviton] (mp) -- [graviton] (x),
	(b) -- [plain] (m) -- [graviton] (mp) -- [graviton] (y),
};
\end{feynman}
\end{tikzpicture}
\end{gathered}	
~~+~
\begin{gathered}
\begin{tikzpicture}[scale=0.6]
\begin{feynman}
\vertex (m) at ( 0, 0);
\vertex (mp) at ( -2, 0);
\vertex (x) at (2,2) {$\ell_1$};
\vertex (y) at (2,-2) {$\ell_2$}; 
\vertex (a) at (-2,2) {$P_1$};
\vertex (b) at (-2,-2) {$P_2$};
\diagram* {
	(a) -- [plain] (m) -- [graviton] (x),
	(b) -- [plain] (m) -- [graviton] (y),
};
\end{feynman}
\end{tikzpicture}
\end{gathered}\nn
\end{equation}
\caption{Tree Diagrams}
\end{figure}
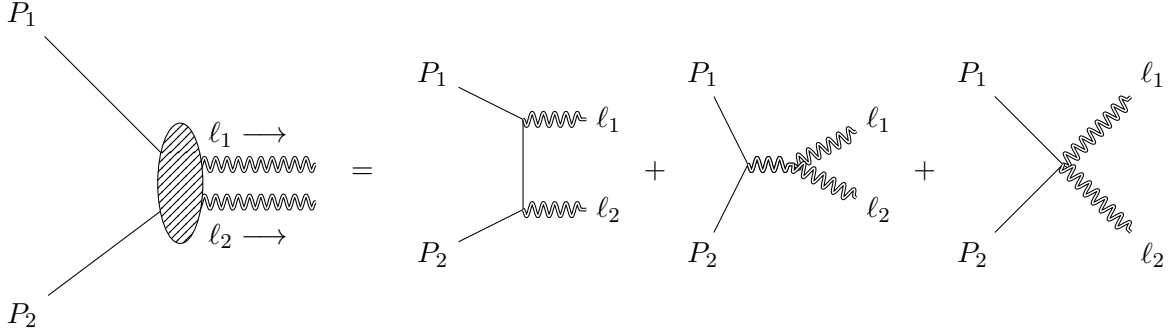

At order $\lambda^1$ in the three-graviton coupling (i.e. ignoring the $\lambda^0$ GR contribution and $\lambda^2$ pure cubic contributions to eq.~\eqref{gravcompton}), and choosing to focus on the $h_1 = h_2 = -$ case, the corresponding amplitudes are given by
\begin{subequations}
\begin{align}\label{key}
M_L[P_1,P_2,\ell_1^{-},\ell_2^{-}]^{(GR)} &= \frac{\kappa^2}{16}\frac{m_A^4\braket{\ell_1\ell_2}^4}{(P_1 + P_2)^2(P_1\cdot\ell_1)(P_2\cdot\ell_1)},\\[1em]
M_R[-\ell_1^+,-\ell_2^+,P_3,P_4]^{(R^3)} &= \frac{3}{16}\frac{\kappa^4\lambda[\ell_1\ell_2]^4}{(P_3+P_4)^2}\Big(\beta_1\left((\ell_1\cdot P_3 - \ell_1\cdot P_4)^2 - m_B^2\ell_1\cdot \ell_2\right)\nn\\ & 
~~~~~~~~~~~~~~~~~~~~~~~~~~ - 8\beta_2(\ell_1\cdot P_3)(\ell_1\cdot P_4)\Big).
\end{align}
\end{subequations}
The loop amplitude is therefore given by\footnote{For the purposes of isolating the contribution of the $G_3$ term, one can express the four point as $$M_R[-\ell_1^+,-\ell_2^+,P_3,P_4]^{(R^3)} = \frac{3}{64}\kappa^4\lambda[\ell_1\ell_2]^4\frac{\beta_1\,t\big(t - 2m_A^2\big) - 16(\beta_1 + 2\beta_2)(\ell_1\cdot P_3)(\ell_1\cdot P_3)}{(P_3+P_4)^2}$$} 
\begin{flalign}\label{eq:2-particle cut loop integral}
M^{(1)}_{4} \ =& \ \frac{3\kappa^{6}\lambda m_{A}^{4}t^{2}}{32}\int\frac{\sd^{4}\ell}{(2\pi)^{4}}\frac{\beta_1\left((\ell\cdot P_3 - \ell\cdot P_4)^2 - \frac{1}{2}m_B^2(P_3+P_4)^2\right) - 8\beta_2(\ell\cdot P_3)(\ell\cdot P_4)}{\ell^{2}(\ell \: - \: P_1 - P_2)^{2}[(\ell \: - \: P_{1})^{2} \: - \: m_{A}^{2}][(\ell \: - \: P_{2})^{2} \: - \: m_{A}^{2}]}.
\end{flalign}
Performing a standard Passarino-Veltman decomposition, we can express the loop amplitude as a sum of boxes, triangles and bubbles. We find that there are no box contributions, but that there are triangles and bubbles. Evaluating these with the help of Package-X \cite{packagex}, we find
\begin{align}\label{loopamp}
M_4^{(1)} &= -\frac{3\kappa^6\lambda m_A^4t^2}{32(t-4m_A^2)^2}\Big[(\beta_1 + 2\beta_2)b_2(t)B_0(t) \nn\\ & ~~~~~~~~~~~~~~~~~~~~~~~~~ + \left[(\beta_1 + 2\beta_2)c_3^{2}(t) +  \beta_1c_3^{1}(t)\right]C_0(P_1^2,P_2^2,t;0,m_A,0)\Big]
\end{align}
where $B_0$ and $C_0$ are the bubble and triangle scalar Passarino-Veltman functions and
\begin{subequations}
\begin{align}\label{key}
b_2(t) &= 6m_A^4 + 4m_A^2\left(m_B^2 -3s\right) + 6(m_B^2 - s)^2 - 2\left(2(m_A^2 + m_B^2) - 3s\right)t + t^2,\\[1em]
c_3^{1}(t) &= \frac{1}{2}(t-4m_A^2)^2(t-2m_B^2),\\[1em]
c_3^{2}(t) &= 2\Big[2m_A^2\left(s-\left(m_A-m_B\right){}^2\right) \left(s-\left(m_A+m_B\right){}^2\right) + \left(- 3m_A^4 + 2m_A^2m_B^2 +(m_B^2 - s)^2 \right)t \nn\\ &~~~~~ + (m_A^2 - m_B^2 + s)t^2\Big].
\end{align}
\end{subequations}
Expanding these terms for small $t$ (and keeping only the terms that give rise to leading order classical and quantum contributions to the potential) we find
\begin{subequations}
\begin{align}\label{key}
\frac{t^2b_2(t)}{(t-4m_A^2)^2} &= \left(3m_A^4 +2m_A^2(m_B^2 -3s) + 3(m_B^2 - s)^2\right)\frac{t^2}{8 m_A^4} +\cl{O}(t^3) %\nn\\ &~~~ + \left(3(m_B^2 - s)^2 - m_A^2(2m_B^2 + m_A^2) \right)\frac{t^3}{16 m_A^6} +\cl{O}(t^4)
,\\[1em]
\frac{t^2c_3^1(t)}{(t-4m_A^2)^2} &= - m_B^2t^2 + \frac12t^3,\\[1em]
\frac{t^2c_3^2(t)}{(t-4m_A^2)^2} &=  \left(s-\left(m_A-m_B\right){}^2\right) \left(s-\left(m_A+m_B\right){}^2\right)\frac{t^2}{m_A^2} + \cl{O}(t^3). %+ \left((m_B^2 - s)^2 - m_A^2(m_A^2 + s)\right)\frac{t^3}{m_A^4} \nn\\ &
%~~~~~+ \cl{O}(t^4).
\end{align}
\end{subequations}
\subsection{Leading Singularity}\label{leadingS}
Computing scattering amplitudes via the unitarity cuts method is an often cumbersome (or impossible) affair, requiring us to solve complicated divergent loop integrals using some regularisation and perhaps a clever method of integrand reduction\footnote{This process can, however, be almost entirely automated nowadays using one of the many excellent available software packages, for example \cite{packagex}.}. Using two-particle cuts, solutions to the cut conditions ensure that the loop momenta remain real and the integrals can be evaluated on those real solutions. However, as is now standard in modern amplitude techniques, considering scattering amplitudes as analytic functions of \textit{complex} momenta often yields incredible simplifications, allowing us to utilise the full barrage of tools bequeathed to us by complex analysis.

In this spirit, we will revisit the calculation of the classical potential in higher derivative gravity by considering the \textit{leading singularity} \cite{ArkaniHamed:2008gz,Britto:2004nc}, the highest codimension singularity of the amplitude found by fully localising every loop integral. In doing so, we find that the solutions to the cut conditions are typically complex and, at one loop, this means that the problem of computing loop amplitudes conveniently reduces to the problem of computing residues of some product of (complex) tree amplitudes. It was recently shown that the leading singularity encodes the information required to compute classical gravitational effects \cite{Cachazo:2017jef,Guevara:2017csg,Guevara:2018wpp,Bautista:2019tdr} and, in this section, we will review the techniques required to compute leading singularities and use them to compute the classical potential in cubic gravity once more, showing that the result is identical to that obtained via unitarity cuts.

Since we require at least one massive propagator in the loop to find classical effects, we consider the triangle diagram
\begin{figure}[H]
	\centering
	\begin{tikzpicture}[scale=1]
	\begin{feynman}
	\vertex (m) at ( -2, 1);
	\vertex (mp) at ( -2, -1);
	\vertex (q) at (0.3, 0);
	\vertex (qp) at (0.3,0);
	\vertex (x) at (-0.1,0);
	\vertex (xp) at (0.1,0);
	\vertex (y) at (-0.1,0);
	\vertex (yp) at (0.1,0);  
	\vertex (a) at (-4,2) {$P_1$};
	\vertex (b) at (-4,-2) {$P_2$};
	\vertex (c) at (2,-2) {$P_3$};
	\vertex (d) at (2,2) {$P_4$};
	\diagram* {
		(a) -- [plain] (m) -- [graviton] (x) (xp) -- [graviton] (q) -- [plain] (d),
		(m) -- [plain] (mp),
		(b) -- [plain] (mp) -- [graviton] (y) (yp) -- [graviton] (qp)  -- [plain] (c)
	};
	%\draw[preaction={fill, white},pattern=north east lines] (-1.9,-0.25) ellipse (0.3cm and 0.8cm);
	\draw[preaction={fill, white},pattern=north east lines] (0,0) ellipse (0.6cm and 0.6cm);
	%\draw [densely dashed, red, line width=0.5mm,] (0,1.2) -- (0,-1.6);
	%\draw (-1,0.2) node[above] {$\ell_1$};
	%\draw (-1,-0.65) node[above] {$\ell_2$};
	\fill[white] (-2,0) ellipse (0.12cm and 0.12cm);
	\fill[white] (-1.15,0.58) ellipse (0.13cm and 0.13cm);
	\fill[white] (-1.165,-0.54) ellipse (0.13cm and 0.13cm);
	\draw [densely dashed, red, line width=0.3mm,] (-0.75,1.25) -- (-1.5,0);
	\draw [densely dashed, red, line width=0.3mm,] (-0.75,-1.25) -- (-1.5,0);
	\draw [densely dashed, red, line width=0.3mm,] (-2.75,0) -- (-1.5,0);
	\fill[white] (-1.48,0) ellipse (0.2cm and 0.2cm);
	%\draw [densely dashed, red, line width=0.5mm,] (-1,0) -- (0,-1.6);
	\end{feynman}
	\end{tikzpicture}
	\caption{LS Triangle Diagram}
\end{figure}
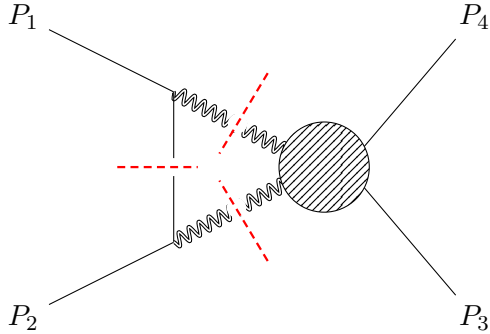
\vspace{1cm}
\vspace{0.2cm}

In order to compute the classical piece of this diagram, we will use the on-shell leading-singularity method presented in \cite{Cachazo:2017jef}. 
%The general procedure is to compute the imaginary part of the amplitude via a maximal cut

To begin with, we will compute the imaginary part of the all-plus contribution to this amplitude, meaning we need to evaluate the integral
\begin{equation}\label{key}
I = \sum_{h_1,h_2}\oint_{\Gamma} \frac{\sd^4L}{(L^2-m^2)k_1^2k_2^2}M_3[P_1,-L,k_1^{-h_1}]M_3[L,P_2,k_2^{-h_2}]M_4[-k_1^{h_1},-k_2^{h_2},P_3,P_4],
\end{equation}
where $k_1 = L + P_1$ and $k_2 = L - P_2$.

We will parameterise the massive loop momenta by
\begin{equation}\label{key}
L = zl + \omega q,
\end{equation}
where $z,\omega \in \mathbb{C}$ are parameters to be integrated over, $l = \lambda\tilde{\lambda}$ is massless and $q$ is an arbitrary fixed reference vector.

Cutting the massive propagator and following \cite{Cachazo:2017jef}, we can write this as
\begin{equation}\label{key}
I = \sum_{h_1,h_2}\oint_{\Gamma_{LS}} \frac{z\,\sd z\braket{\lambda \,\sd\lambda}[\tilde{\lambda} \,\sd\tilde{\lambda}]}{k_1^2k_2^2}M_3[P_1,-L,k_1^{-h_1}]M_3[L,P_2,k_2^{-h_2}]M_4[-k_1^{h_1},-k_2^{h_2},P_3,P_4],
\end{equation}
We can also project the external momentum onto the lightcone using massless vectors $p_1 = \lambda_1\tilde{\lambda}_1$ and $p_2 = \lambda_2\tilde{\lambda}_2$
\begin{equation}\label{paramp3p4}
P_1 = p_1 + xp_2,~~~~~P_2 = p_2 + xp_1,~~~~~ x = \frac{m_A^2}{2p_1\cdot p_2},
\end{equation}
where we have used $P_1^2 = P_2^2 = m_A^2$ to fix $x$. We note that, since we are going to look primarily at the $t$-channel, we can use $x$ to to parameterize it as
\begin{equation}\label{key}
\frac{(1+x)^2}{x} = \frac{t}{m_A^2},~~~~~\frac{(1-x)^2}{x} = \frac{t - 4m_A^2}{m_A^2}.
\end{equation}
If we now also choose two mixed reference vectors $q = \lambda_1\tilde{\lambda}_2$ and $\bar{q} = \lambda_2\tilde{\lambda}_1$, then we have four linearly independent vectors which we can use as a basis for our massless loop amplitude, i.e.
\begin{equation}\label{key}
l = Ap_1 + Bp_2 + Cq + \bar{q}.
\end{equation}
Demanding the on-shell condition $l^2 = 0$ gives $C = AB$, and regarding $A,B \in \mathbb{C}$ means we can identify $\sd A\,\sd B \propto \braket{\lambda\, \sd\lambda}[\tilde{\lambda} \,\sd\tilde{\lambda}]$. After a change of variables, we find
\begin{align}\label{key}
I &= \frac{1}{(2\pi i)^3}\frac{(2p_1\cdot p_2)}{16}\sum_{h_1,h_2}\oint_{\Gamma_{LS}}\frac{z\,\sd z\,\sd A\,\sd B~~M_4[P_3,P_4,k_1^{h_1}k_2^{h_2}] M_3[k_2^{-h_2},P_2,L] M_3[k_1^{-h_1},P_1,-L]}{(m_A^2+z(p_1\cdot p_2)(B+xA))(-m_A^2+z(p_1\cdot p_2)(A+xB))}
\end{align}
Using eq.~\eqref{paramp3p4}, we find poles at $A = -B = \frac{2x}{z(1-x)}$, leaving finally after integration over $A,B$
\begin{equation}\label{keyint}
I = \frac{x}{4m_A^2(1-x^2)}\sum_{h_1,h_2}\frac{1}{2\pi i}\oint_{\Gamma_{LS}}\frac{\sd z}{z}M_3[P_1,-L,k_1^{-h_1}]M_3[L,P_2,k_2^{-h_2}]M_4[-k_1^{h_1},-k_2^{h_2},P_3,P_4].
\end{equation}
With this in hand, we can now parameterize $k_3$ and $k_4$ using the same parameters. For $k_3$, this is
\begin{align}\label{k3Param}
k_1 &= L + P_1 \nn\\
&= (1+zA)p_1 + (zB+x)p_2 + (zC + \omega)q + z\bar{q}\nn \\
&= r(x)p_1 - xr(x) p_2 - \frac{xr^2}{z}q + z\bar{q}\nn \\
&= r(x)\Big[\lambda_1 + \frac{z}{r(x)}\lambda_2\Big]\Big[\tilde{\lambda}_2 - \frac{x}{z}r(x)\tilde{\lambda}_1\Big],
\end{align}
where we have defined \smash{$r(x) = \frac{1+x}{1-x}=\left(\frac{t}{t-4m_A^2}\right)^{1/2}$} and plugged in $\omega = \frac{m_A^2}{2z(q\cdot\bar{q})} = -\frac{x}{z}$. Repeating this for $k_4$, we find
\begin{equation}\label{key}
k_2 = r(x)\Big[\lambda_2 + \frac{x}{z}r(x)\lambda_1\Big]\Big[\tilde{\lambda}_2 - \frac{z}{r(x)}\tilde{\lambda}_1\Big].
\end{equation}
%\subsubsection{Tree-Level Amplitudes}
It follows from these parameterizations, that 
\begin{equation}\label{param dot products}
k_1\cdot P_i = \frac{1}{z}\left[z^2\bar{q}\cdot P_i + r(x)z(p_1 - xp_2)\cdot P_i - xr^2(x)q\cdot P_i \right], 
\end{equation}
for $i=3,4$. Ultimately, we want to express everything in terms of Mandelstam invariants. Using eq.~\eqref{paramp3p4}, we find that we can write
\begin{align}\label{key}
(p_1 - xp_2)\cdot P_3 &= \frac12\left(\frac{1+x}{1-x}\right)\left(m_A^2-m_B^2+s\right),\\
(p_1 - xp_2)\cdot P_4 &= \frac12\left(\frac{1+x}{1-x}\right)\left(m_A^2-m_B^2+u\right).
\end{align}
As such, eq.~\eqref{param dot products} becomes
\begin{subequations}
	\begin{align}
	k_1\cdot P_3 &= \frac{1}{z}\left[z^2\bar{q}\cdot P_3 + \frac12 r^2(x)\left(m_A^2-m_B^2+s\right)z - xr^2(x)q\cdot P_3 \right], \\[1em] k_1\cdot P_4 &= \frac{1}{z}\left[z^2\bar{q}\cdot P_4 + \frac12 r^2(x)\left(m_A^2-m_B^2+u\right)z - xr^2(x)q\cdot P_4 \right].   
	\end{align}
\end{subequations}
Moreover, as $z$ is our integration variable, we are free to make the following rescaling: $z\to 2\frac{1+x}{M^2}\sqrt{-x}(q\cdot P_3)z$, where $M$ is defined by $M^4 \coloneqq -4(1-x)^2(q\cdot P_3\,\bar{q}\cdot P_3) = (m_A^2 - m_B^2)^2 - su$ (note that \smash{$(q\cdot P_3)= -(q\cdot P_4)$}, and \smash{$(q\cdot P_3\,\bar{q}\cdot P_3)=(q\cdot P_4\,\bar{q}\cdot P_4)$}). From this, we find that
\begin{subequations}\label{k1P3 k1P4}
	\begin{align}
	k_1\cdot P_3 &= \frac{M^2m_Ar^2(x)}{2\sqrt{-t}}\frac{1}{z}\left[-z^2 - \left(m_B^2-m_A^2-s\right)\frac{\sqrt{-t}}{M^2m_A}z + 1 \right] \coloneqq \frac{M^2m_Ar^2(x)}{2\sqrt{-t}}\frac{1}{z}\,F_s(z), \\[1em] k_1\cdot P_4 &= \frac{M^2m_Ar^2(x)}{2\sqrt{-t}}\frac{1}{z}\left[z^2 - \left(m_B^2-m_A^2-u\right)\frac{\sqrt{-t}}{M^2m_A}z - 1 \right] \coloneqq \frac{M^2m_Ar^2(x)}{2\sqrt{-t}}\frac{1}{z}\,F_u(z) ,  
	\end{align}
\end{subequations}
where $F_s$ and $F_u$ are defined as 
\begin{subequations}
	\begin{align}\label{Fs Fu}
	F_s(z) =& -z^2 -\frac{\sqrt{-t} \left(m_B^2-m_A^2-s\right)}{m_A \sqrt{-2 m_A^2 \left(m_B^2+s\right)+m_A^4-2 s m_B^2+m_B^4+s (s+t)}}z + 1 ,\\[1em] F_u(z) =& \ z^2 -\frac{\sqrt{-t} \left(m_B^2-m_A^2-u\right)}{m_A \sqrt{-2 m_A^2 \left(m_B^2+s\right)+m_A^4-2 s m_B^2+m_B^4+s (s+t)}}z - 1.
	\end{align}
\end{subequations}
We now need to compute the tree-level amplitudes with which we will build the loop. These are give as follows:
\begin{subequations}
	\begin{align}\label{key}
	& M_4[P_3,P_4,k_1^+,k_2^+]^{(R^3)} = \frac{3}{64}\kappa^4\lambda[k_1k_2]^4\frac{\beta_1\,t\big(t - 2m_A^2\big) - 16(\beta_1 + 2\beta_2)(k_1\cdot P_3)(k_2\cdot P_4)}{(P_3+P_4)^2}\,,\\[1em]
	&M_4[P_3,P_4,k_1^+,k_2^-]^{(GR)} = \frac{\kappa^2}{4}\frac{[k_1|P_3\ket{k_2}^2}{(P_3+P_4)^2[k_1|P_3\ket{k_1}[k_1|P_4\ket{k_1}},
	\end{align}
\end{subequations}
and,
\begin{equation}\label{key}
M_3[P_1,P_2,k^+]^{(GR)} = \frac{\kappa}{2}\frac{\bra{g}P_1|k]^2}{\braket{gk}^2},~~~~~M_3[P_1,P_2,k^-]^{(GR)} = \frac{\kappa}{2}\frac{[g|P_1\ket{k}^2}{[gk]^2}.
\end{equation}
Note that the contributions from cubic gravity (see Appendix B for the full derivation) to the four-point amplitudes arise only in the all-positive (and all-negative) helicity case. The reason being is that cubic gravity only affects the three-point vertex function for graviton self-interactions when each graviton has the same helicity (all positive, or all negative).

Focusing on pieces with only one cubic gravity contribution to the loop, and making use of eq.~\eqref{k1P3 k1P4}, one can recast $M_4[P_3,P_4,k_1^+,k_2^+]^{(R^3)}$ into the following form:

\begin{align}
M_4[P_3,P_4,k_1^+,k_2^+]^{(R^3)} &= \frac{3}{64}\kappa^4\lambda [k_1k_2]^4\,\beta_1(t-2m_B^2) \nn\\[0.4em] &\quad +\frac{3}{16t^2}\kappa^4\lambda M^4m_A^2r^4(x)[k_1k_2]^4\,\frac{1}{z^2}\,(\beta_1 + 2\beta_2)\,F_s(z)\,F_u(z).
\end{align}
To evaluate the leading singularity, we need to plug these tree level amplitudes into eq. \eqref{keyint}, with $h_1 = h_2 = +$, and integrate over the localised integral, i.e. take residues. For this we need to include the product of three-points, given by
\begin{align}
M_3[k_2^-,P_2,L]^{(GR)}M_3[k_1^-,P_1,-L]^{(GR)} &= \frac{\kappa^{2}}{4}\frac{[p_1|P_2\ket{k_2}^2}{[p_1k_2]^2}\frac{[p_2|P_1\ket{k_1}^2}{[p_2k_1]^2} = \frac{\kappa^{2}}{4}[p_1p_2]^4\frac{\braket{p_2k_2}^2\braket{p_1k_1}^2}{[p_1k_2]^2[p_2k_1]^2}\nn\\ &= \frac{\kappa^{2}}{4}x^2r^4(x)\braket{p_1p_2}^4.
\end{align}
where we have chosen the reference vectors in the three-points to be $p_1$ and $p_2$, and we have made use of eqs.~\eqref{paramp3p4},~\eqref{k3Param}, and the relations: $[p_1k_{2}]=[p_1p_2]$, $[p_1k_{2}]=\frac{xr(x)}{z}[p_1p_2]$, $\braket{p_1k_{1}}=z\braket{p_1p_2}$ and $\braket{p_2k_{2}}=-\frac{xr^{2(x)}}{z}\braket{p_1p_2}$.

We will consider the contribution common to both $\beta_1$ and $\beta_2$ first which, after using definitions of $r(x)$ and $M^4$, is given by (using that $[k_1k_2]=(1-x)[p_1p_2]$)

\begin{align}\label{key}
I_{++}^{(\beta_1+2\beta_2)} &= \frac{3\kappa^6\lambda m_A^6 M^4}{256t}\left(\frac{t}{t-4m_A^2}\right)^{5/2} \frac{1}{2\pi i}\oint_{\Gamma_{LS}}\frac{\sd z}{z^3}~ F_s(z) ~F_u(z)\nn\\
&= \frac{3\kappa^6\lambda m_A^4}{256}\left(\frac{t}{t-4m_A^2}\right)^{5/2}\left[tR(s,m_A,m_B) + P(s,m_A,m_B) + \frac{Q(s,m_A,m_B)}{t}\right]\nn\\[1em] &= I^{(\beta_1+2\beta_2)}_{--}\,,
\end{align}
where
\begin{subequations}
	\begin{align}
	R(s,m_A,m_B) &= m_A^2-m_B^2+s,\\[1em]
	P(s,m_A,m_B) &= 2m_A^2 m_B^2+\left(s-m_B^2\right)^2-3 m_A^4,\\[1em]
	Q(s,m_A,m_B) &= 2m_A^2\left(s-\left(m_A+m_B\right)^2\right) \left(s-\left(m_A-m_B\right)^2\right).
	\end{align}
\end{subequations}
In anticipation of comparison with the unitary cuts calculation, we observe that $$c^2_3(t)= 2\big[tR(s,m_A,m_B) + P(s,m_A,m_B) + \frac{Q(s,m_A,m_B)}{t}\big],$$ and thus $$I_{++}^{(\beta_1+2\beta_2)}=-\frac{3\kappa^6\lambda m_A^4}{512}\left(\frac{t}{t-4m_A^2}\right)^{5/2}c^2_3(t).$$

The remaining $\beta_1$ terms are evaluated analogously, finding
\begin{align}
I_{++}^{\beta_1} &= \frac{3\kappa^6\lambda m_A^6}{1024}r^3(x)\frac{(1-x)^2}{x}(t-2m_A^2)\nn\\
&=\frac{3\kappa^6\lambda m_A^4}{1024t }\left(\frac{t}{t-4m_A^2}\right)^{5/2}(t - 4m_A^2)^2(t - 2m_B^2)\nn\\[1em] &= I_{--}^{\beta_1}\,.
\end{align}
Here, we note that $c^1_3(t)=\frac{1}{2}(t - 4m_A^2)^2(t - 2m_B^2)$, such that $I_{++}^{\beta_1}=-\frac{3\kappa^6\lambda m_A^4}{512t }\left(\frac{t}{t-4m_A^2}\right)^{5/2}c^1_3(t)$.

In general, the full amplitude is related to the imaginary part $I$ by the dispersion relation of a given channel: we integrate the imaginary part of the amplitude along the branch cut in order to reconstruct the entire amplitude \cite{Donoghue_1996}. In this case, as was shown in \cite{Cachazo:2017jef}, the amplitude has a \textit{double} discontinuity in the $t$-channel, and we must integrate along both, meaning the full amplitude is given by the dispersion relation 
\begin{equation}\label{key}
M^{(1)}(s,t)_{++} = \frac{1}{2\pi i}\int_{0}^{4m_A^2}\frac{1}{2\pi i}\int_{0}^{4m_A^2}\frac{\sd t''}{t'-t''}\frac{\sd t'}{t-t'} I_{++}(s,t').
\end{equation}
We need to integrate three integrands: $J$, $tJ$ and $J/t$, where
\begin{equation}\label{key}
J(t) = \left(\frac{t}{t-4m_A^2 + \epsilon}\right)^{5/2},
\end{equation}
and we have included the $\epsilon$ in order to regulate the divergence. Integrating this (and neglecting terms that vanish in the $\epsilon\to 0$ limit), we find
\begin{equation}\label{key}
\int_0^{4m_A^2}\frac{\sd t'}{t-t'}\, J(t') = 2\tanh ^{-1}\left(2 \sqrt{\frac{m_A^2}{\epsilon}}J^{-1/5}(t)\right)J(t) - \log \left(\frac{(2m_A+\sqrt{\epsilon })^2}{\epsilon -4 m_A^2}\right) + \cl{O}\left(\frac{1}{\epsilon^{5/2}}\right).
\end{equation}
Keeping only the parts finite in $\epsilon$ and taking $\epsilon \to 0$, we find
\begin{equation}\label{key}
\int_0^{4m_A^2} \frac{\sd t'}{t-t'}\, J(t') = -i\pi J(t) + i\pi,
\end{equation}
where we can ignore the addition of $i\pi$ knowing that it comes from the log pieces\footnote{These only contribute to the quantum piece of the amplitude \cite{Holstein_Ross_2008}.}. This means that our integral is \textit{classically} self-similar, and all of the integrations we need to do are therefore trivial, meaning we can write
\begin{equation}\label{key}
M^{(1)}(s,t) = \frac14 \sum_{\beta_1,\beta_2}\sum_{h_1=h_2}I_{h_1h_2}(s,t) + m_A \leftrightarrow m_B.
\end{equation}
Evaluating this, we find that the leading singularity is
\begin{align}\label{eq:LS final result}
M^{(1)}_4 &= -\frac{3\kappa^6\lambda m_A^4}{1024t}\left(\frac{t}{t-4m_A^2}\right)^{5/2}\bigg[(\beta_1 + 2\beta_2)c_3^2(t) + \beta_1c_3^1(t)\bigg].
\end{align}
We see then that the structure of the leading singularity is identical to the classical piece found by directly computing the loop via unitarity cuts. In fact, when evaluating only the classical part of the finite PV integrals, we find that it is given by
\begin{align}\label{key}
C_0(m_A^2,m_A^2,t;0,m_A,0)_{classical} &= \frac{\pi^2 + 3\Li_2\left(1 + \left(\sqrt{1-\frac{4 m_A^2}{t}}-1\right)\frac{t}{2 m_A^2}\right)}{24\pi^2t}\sqrt{\frac{t}{t-4m_A^2}}\\
&\simeq\frac{1}{16t}\sqrt{\frac{t}{t-4m_A^2}},
\end{align}
where we have kept only the first term in the expansion of $\Li_2(1 + f[t]) \simeq \frac{\pi^2}{6}$.

Plugging this into the unitarity cuts computation and ignoring the quantum corrections show that these match exactly.

Before proceeding, we make a brief comment about extracting the classical contributions to the amplitude~\eqref{eq:LS final result}. This can be achieved by appealing to the holomorphic classical limit (HCL) of the amplitude derived via the leading singularity approach (see ref.~\cite{Guevara:2017csg} for a detailed analysis of the procedure). The HCL corresponds to taking the limit $x\to -1$ (i.e. $t/m_A^2 \to 0$), which in practice means that we should retain only the leading-order-in-$t$ contributions in eq.~\eqref{eq:LS final result}. Terms proportional to the same order in $t$ but multiplied by $\log\left(-\frac{t}{m^2}\right)$ will therefore be $\propto \hbar$. This is equivalent to the method of restoring factors of $\hbar$ to extricate the classical and quantum components (see~\cite{Kosower:2018adc} for further details on this approach). 
\section{Potential and Black Hole Solutions}
With the one-loop amplitude in hand, we can now go about deriving the potential. In order to derive the non-relativistic limit, we evaluate the amplitude~\eqref{loopamp}, taking the small $t$ limit of the (finite) integrals and summing together with $m_A\leftrightarrow m_B$. We will focus on the unitarity cuts calculation since this also gives all of the quantum corrections. In the small-$t$ limit, the PV integrals are given by
\begin{subequations}
\begin{align}
    B_0(t) &\simeq \frac{1}{16\pi^2}\log(-t),\\[1em]
    C_0(m_A^2,m_A^2,t;0,m_A,0) &\simeq \frac{1}{32\pi^2m_A^2}\left[\log\left(\frac{m_A^2}{t}\right) +\frac{\pi^2m_A}{\sqrt{-t}}\right], 
\end{align}
\end{subequations}
Consequently we derive the following amplitude for small $t$ (up to $\cl{O}(t^{2})$)
\begin{align}
    M_4^{(1)} =& -\frac{3}{4096} \kappa^6\lambda (\beta_1+2\beta_2)\left(m_A + m_B\right)\Big[\left(\left(m_A-m_B\right){}^2-s\right) \left(\left(m_A+m_B\right){}^2-s\right)(-t)^{3/2}\Big]\nn \\[0.5em]& + \frac{3}{1024}\kappa ^6\lambda\beta_1\Big[m_A^2m_B^2(m_A + m_B)(-t)^{3/2}\Big]\nn\\[0.5em]
    & -\frac{3\hbar}{512\pi^2}\kappa^6\lambda(\beta_1+2\beta_2)\left((m_A^2-s)^2 + (m_B^2-s)^2 - s^2\right)t^2\log\left(-t\right)\nn\\[0.5em]
    & + \frac{3\hbar}{512\pi^2}\kappa^6\lambda\beta_1 m_A^2m_B^2t^2\log(-t). 
\end{align}

We can take the fully non-relativistic limit of this via\footnote{We thank the authors of \cite{Brandhuber:2019qpg} for very useful discussions on this point}
\begin{subequations}
\begin{align}\label{key}
t &\longrightarrow -\mathbf{q}^2,\\
s &\longrightarrow (m_A+m_B)^2 \left(1 + \frac{\textbf{p}^2 + \frac14 \textbf{q}^2}{m_Am_B} \right),
\end{align}
\end{subequations}
which leaves us with a momentum space potential
\begin{align}\label{key}
V(\mathbf{q},\mathbf{p}) &= V_\text{cl}(\mathbf{q},\mathbf{p}) + \hbar V_\text{qu}(\mathbf{q},\mathbf{p}), 
\end{align}
where $V_\text{cl}(\mathbf{q})$ and $V_\text{qu}(\mathbf{q})$ are the classical and quantum contributions, respectively given by
\begin{subequations}
 \begin{align}
   V_\text{cl}(\mathbf{q},\mathbf{p}) &= -\frac{3}{4096}\kappa^6\lambda(\beta_1 + 2\beta_2)\bigg[\frac{(m_A+m_B)^3}{m_Am_B}\mathbf{p}^2|\mathbf{q}|^3\bigg] \nn\\[0.5em] &~+ \frac{3}{8192}\kappa^6\lambda\beta_1\frac{(m_A + m_B)}{m_Am_B}\bigg[2m_A^2m_B^2|\mathbf{q}|^3 - (m_A^2+m_B^2)\mathbf{p}^2|\mathbf{q}|^3\bigg] + \cl{O}(|\mathbf{q}|^5)\\[1em]
   %
   %
   % Quantum Piece
   %
   %
   V_\text{qu}(\mathbf{q},\mathbf{p}) &= -\frac{3}{2048\pi^2}\kappa^6\lambda\frac{(\beta_1+2\beta_2}{m_Am_B}\bigg[2m_A^2m_B^2 + (3m_A^2 + 8m_Am_B+3m_B^2)\mathbf{p}^2\bigg]\mathbf{q}^4\log\left(\mathbf{q}^2\right)\nn\\[0.5em]
  &~+ \frac{3}{4096\pi^2}\kappa^6\lambda\beta_1\bigg[2m_Am_B - \frac{(m_A^2 + m_B^2)}{m_Am_B}\mathbf{p}^2\bigg]\mathbf{q}^4\log(\mathbf{q}^2) +\cl{O}(|\mathbf{q}|^6).
 \end{align}
\end{subequations}
Taking the Fourier transform as in appendix \ref{appendix:FT}, we find
\begin{equation}\label{finalpotential}
V(\mathbf{r},\mathbf{p}) = V_\text{cl}(\mathbf{r},\mathbf{p}) +\hbar V_\text{qu}(\mathbf{r},\mathbf{p}),
\end{equation}
where
\begin{subequations}
 \begin{align}
  V_\text{cl}(\mathbf{r},\mathbf{p}) &= \frac{9}{1024\pi^2}\kappa^6\lambda(\beta_1 + 2\beta_2) \bigg[\frac{(m_A+m_B)^3}{m_Am_B}\frac{\mathbf{p}^2}{r^6}\bigg] \nn\\[0.5em] &~~- \frac{9}{2048\pi^2}\kappa^6\lambda\beta_1(m_A + m_B)\bigg[2\frac{m_Am_B}{r^6} - \frac{(m_A^2+m_B^2)}{m_Am_B}\frac{\mathbf{p}^2}{r^6}\bigg]  +\cl{O}(r^{-8}),\\[1em]
  %
  %
  % Quantum
  %
  %
  V_\text{qu} (\mathbf{r},\mathbf{p}) &= -\frac{45}{512\pi^3}\kappa^6\lambda(\beta_1+2\beta_2)\bigg[\frac{2m_Am_B}{r^7} + \frac{(3m_A^2 + 8m_Am_B + 3m_B^2)}{m_Am_B}\frac{\mathbf{p}^2}{r^7}\bigg] \nn\\[0.5em] &~~~  +\frac{45}{1024\pi^3}\kappa^6\lambda\beta_1\bigg[2\frac{m_Am_B}{r^7} - \frac{(m_A^2 + m_B^2)}{m_Am_B}\frac{\mathbf{p}^2}{r^7}\bigg] +\cl{O}(r^{-9}).
 \end{align}
\end{subequations}
%+ \cl{O}(r^{-10})
Equipped with the expressions for the classical and quantum corrections to the potential, we can make contact with some specific theories my making particular choices for the couplings. Firstly, we derive the first order corrections to the potential arising from ECG, by choosing $\beta_1 = 12, \beta_2 = 1$ and restoring $G$ via $\kappa = \sqrt{32\pi G}$. Doing so gives the potential
\begin{equation}
    V_{ECG}(\textbf{r}) = 72\tilde{\lambda}G^4m_Am_B\left(\frac{3(m_A+m_B)}{r^6} -\hbar\frac{100}{\pi r^7}\right).
\end{equation}
where we have rescaled the coupling by $$\lambda\to -\frac{G\tilde{\lambda}}{16\pi}\,,$$
in order to ensure that our $\lambda$ matches the one in ref. \cite{Bueno:2016xff}.

We can also compute the corrections to the potential that arise from the $\alpha'^2$ part of low-energy effective action in string theory, recently discussed in~\cite{Brandhuber:2019qpg}, which enter as
\begin{equation}
    S = -\frac{2\alpha'^2}{\kappa^2}\int d^4x \sqrt{-g}\left(\frac{1}{48}I_1 + \frac{1}{24}G_3\right),
\end{equation}
where $I_1$ is the $\beta_2$ term in our case. To compute the $I_1$ correction, we take $\lambda = -\frac{2\alpha'^2}{\kappa^2}$ and then $\beta_1 = 0,~\beta_2 = \frac{1}{48}$ which gives
\begin{equation}
    V_{I_1}(\textbf{r}) = 3(\alpha'G)^2\left(\frac{(m_A+m_B)^3}{4m_Am_B}\frac{\textbf{p}^2}{r^6} - \hbar\frac{5m_Am_B}{\pi r^7}\right).
\end{equation}
Then to compute the $G_3$ contribution we take $\beta_1 = -2\beta_2 = -\frac{1}{24}$ to find
\begin{equation}
    V_{G_3}(\textbf{r}) = \frac{3(\alpha'G)^2}{4}m_Am_B\left(\frac{(m_A+m_B)}{r^6} - \hbar\frac{10}{\pi r^7}\right).
\end{equation}

Furthermore, given the potential we can, as discussed in section \ref{potentials} above, derive a static, spherically symmetric black hole solution. Knowing the form of the Schwarzschild solution, and noting the argument earlier that any solution derived from a cubic theory must be a correction to this, we find a black hole solution of the form
\begin{equation}
f(\mathbf{r}) = 1 - \frac{2Gm_A}{r} - 36\beta_1\frac{G^4\tilde{\lambda} m_A^2}{r^6} + 360\hbar\left(3\beta_1 +4\beta_2\right)\frac{G^4\tilde{\lambda}m_A}{\pi r^7}.
\end{equation}
Choosing the specific coefficients in ECG, we find a solution of the form
\begin{equation}
f(\mathbf{r}) = 1 - \frac{2Gm_A}{r} - 432\frac{G^4\tilde{\lambda} m_A^2}{r^6} + 14400\hbar\frac{G^4\tilde{\lambda}m_A}{\pi r^7}\,.
\end{equation}
The classical part of this metric matches those derived from Einsteinian cubic gravity \cite{Hennigar:2016gkm,Bueno:2016lrh}. 

The solutions found in those papers were not easy to come by, being the (perturbative) solution to a particularly complicated differential equation with apparently no analytic solution. Here, we have come to the same solution by considering gravity as a quantum field theory and using the tools of modern scattering amplitudes, deriving the quantum corrections to the metric as an added bonus. Furthermore, we showed how the same \textit{classical} black hole solution could be obtained by computing residues of the leading singularity.

\section{Discussion}
In this paper, we have studied the leading order dynamics of a general cubic theory of gravity coupled to a spin-zero matter field, within the framework of the modern scattering amplitude techniques. 

We observed that the effects of cubic gravity on the purely classical graviton mediated interaction between two scalars can only occur at one-loop order and above, at least when considering minimal coupling. Loops can provide classical contributions~\cite{Holstein:2004dn, Kosower:2018adc} and, moreover, any classical contributions to the gravitational potential manifest in diagrams containing massive propagators. Given this, we computed the double-cut of the appropriate one-loop diagram, in which we cut two internal graviton lines and retained only the leading order contributions to the result. 

We then repeated this calculation using an alternative approach: computing the leading singularity of the one-loop triangle diagram with one massive propagator. Not only was this a much more straightforward calculation but also, importantly, we recovered the result we found using the standard unitarity cuts method. This shows that our result is consistent. 

Whilst this draft was in preparation, Brandhuber and Travaglini published work in which they consider the (dynamically) non-trivial cubic corrections to the gravitational action arising from string theory~\cite{Brandhuber:2019qpg}. When comparing results, we find that the corresponding cubic modification of the gravitational potential agrees with theirs, up to an additional non-dispersive contribution. An interesting difference between our results is this additional contribution since its structure is such that, in an appropriate probe-limit, we were able to derive a black hole solution which exactly corresponds to the leading order Einsteinian cubic gravity contribution, matching the result found in ref.~\cite{Bueno:2016lrh}. 

It is interesting to note that this black hole solution survives the limit Einstein gravity plus a pure cubic Gauss-Bonnet term, $G_3$, and that the black hole solution arises from the non-minimal coupling between the spin-zero matter field and the Gauss-Bonnet combination. This is not wholly unexpected since, although possessing trivial dynamics in isolation, it has been shown that the Gauss-Bonnet combination has non-trivial effects on four-point amplitudes when coupled to the matter sector~\cite{1987PhLB..185...52M,Broedel:2012rc}.
\section{Acknowledgements}
We would like to thank Daniel Burger, Ra\'ul Carballo-Rubio, Alfredo Guevara, Jeff Murugan, and Amanda Weltman for useful discussions during the preparation of this work. WTE would like to thank Amanda in particular for hosting him at UCT for the majority of this project. In addition, WTE acknowledges support from the South African Research Chairs Initiative for supporting this visit. NM would like to thank Imperial College London and Nottingham University for hosting him while some of this work was completed. NM is supported by the South African Research Chairs Initiative of the Department of Science and Technology and the National Research Foundation of South Africa. Any opinion, finding and conclusion or recommendation expressed in this material is that of the authors and the NRF does not accept any liability in this regard. 
% % Everything goes above here
\newpage
\section*{Appendix}
\appendix
\section{Tree-Level 4-point Amplitude}\label{appendix:4pt}
To derive the current needed to compute amplitudes involving cubic invariants, we expand the following around flat space (using xAct \cite{xAct})
\begin{equation}\label{key}
\mathcal{P}=\sqrt{-g}\left[\beta_1 R_{a\ b}^{\ c \ d}R_{c\ d}^{\ e \ f}R_{e\ f}^{\ a \ b}+\beta_2 R_{ab}^{cd}R_{cd}^{ef}R_{ef}^{ab} + \beta_3 R_{abcd}R^{ac}R^{bd} + \beta_4 R_{a}^{b}R_{b}^{c}R_{c}^{a}\right].
\end{equation}

Expanding to cubic order and including a de Donder gauge-fixing term, we find that only the terms with coefficient $\beta_1$ and $\beta_2$ survive the requirements that the polarization tensors be transverse (i.e. that $k_\mu \epsilon^\mu(k) = 0$) and the on-shell condition $k^2 = 0$. Since terms of cubic order only contribute same-helicity 3-points, we put two of the legs of the 3-point on shell and fix their helicity to both be identical. Factoring out the third leg, we derive the following currents
\begin{align}\label{negcurrent}
J^{\mu\nu}_{--,--} &= \kappa^3\lambda\frac{3}{16}\braket{12}^4\Big[\beta_1\big(k_1^\mu k_1^\nu + k_2^\mu k_2^\nu - k_1^\mu k_2^\nu - k_2^\mu k_1^\nu\big) \nn\\ & ~~~~~~~~~~~~~~~~~~~~ - \beta_2\big(\bra{1}\gamma^\mu|2]\bra{2}\gamma^\nu|1] + \bra{1}\gamma^\nu|2]\bra{2}\gamma^\mu|1]\big)\Big],\\[0.8em]
J^{\mu\nu}_{++,++} &= \kappa^3\lambda\frac{3}{16}[12]^4\Big[\beta_1\big(k_1^\mu k_1^\nu + k_2^\mu k_2^\nu - k_1^\mu k_2^\nu - k_2^\mu k_1^\nu\big) \nn\\ & ~~~~~~~~~~~~~~~~~~ - \beta_2\big(\bra{2}\gamma^\mu|1]\bra{1}\gamma^\nu|2] + \bra{2}\gamma^\nu|1]\bra{1}\gamma^\mu|2]\big)\Big].
\end{align}
The scalar-scalar-graviton vertex, and the graviton propagator (in the de Donder gauge) are further given by, respectively
\begin{equation}
 J_{00}^{\mu\nu} = -\frac{\kappa}{2}\big[P_3^{\mu}P_4^{\nu}+P_3^{\nu}P_4^{\mu}-\eta^{\mu\nu}\left(P_3\cdot P_4 +m^2\right)\big],
\end{equation}
and
\begin{equation}
 P_{\mu_1\nu_1;\mu_2\nu_2} =-\frac{1}{2k^2}\big[\eta_{\mu_1\mu_2}\eta_{\nu_1\nu_2}+\eta_{\mu_1\nu_2}\eta_{\nu_1\mu_2}-\eta_{\mu_1\nu_1}\eta_{\mu_2\nu_2}\big].\\[0.8em]
\end{equation}
Contracting these with $J^{\mu\nu}_{--,--}$ (and $J^{\mu\nu}_{00}$), we derive the 4-points
\begin{align}\label{key}
M_4[k_1^-,k_2^-,P_3,P_4] &= \frac{3}{16}\kappa^4\lambda\braket{12}^4\frac{\beta_1\left((k_2\cdot P_3 - k_2\cdot P_4)^2 - m^2k_1\cdot k_2\right) - 8\beta_2(k_1\cdot P_3)(k_2\cdot P_3)}{(P_3+P_4)^2}\\[1em]
M_4[k_1^+,k_2^+,P_3,P_4] &= \frac{3}{16}\kappa^4\lambda[12]^4\frac{\beta_1\left((k_2\cdot P_3 - k_2\cdot P_4)^2 - m^2k_1\cdot k_2\right) - 8\beta_2(k_1\cdot P_3)(k_2\cdot P_3)}{(P_3+P_4)^2}
\end{align}
where, in the case of ECG, $\beta_1 = 12$ and $\beta_2 = 1$. Aside from this, if we take $\beta_1 = -2\beta_2$ then the 4-point amplitudes above remain non-trivial, induced by a cubic Gauss-Bonnet $G_3$-interaction. In this case, they reduce to
\begin{align}\label{key}
\frac{1}{\lambda}M_4[k_1^-,k_2^-,P_3,P_4]^{(G_3)} &= \frac{3!}{4}\Big(\frac{\kappa}{2}\Big)^4\braket{12}^4\big(t+2m^2\big)\;,\\[1em]
\frac{1}{\lambda}M_4[k_1^+,k_2^+,P_3,P_4]^{(G_3)} &= \frac{3!}{4}\Big(\frac{\kappa}{2}\Big)^4[12]^4\big(t+2m^2\big)\;,
\end{align}
in agreement with the result found in~\cite{Brandhuber:2019qpg}.
\section{Fourier Transforms}\label{appendix:FT}
We need to compute the Fourier transform of $|\mathbf{q}|^n$, where $n$ is positive
\begin{equation}\label{key}
F[\mathbf{r},n] = \int \frac{\sd^3\mathbf{q}}{(2\pi)^3}e^{i\mathbf{q}\cdot\mathbf{r}}|\mathbf{q}|^n.
\end{equation} 

Formally, this diverges and requires regularization. To do so, we shift $\mathbf{q}$ by a regulator $\epsilon$, i.e.
\begin{equation}\label{key}
F[\mathbf{r},n] = \int \frac{\sd^3\mathbf{q}}{(2\pi)^3}e^{i\mathbf{q}\cdot\mathbf{r} - \epsilon|\mathbf{q}|r}|\mathbf{q}|^n,
\end{equation}
where $\epsilon \ll 1$ and we discard higher orders. Switching to spherical-polar coordinates, this becomes
\begin{equation}\label{key}
F[\mathbf{r},n] = \int_{0}^\pi \sd\theta\int \frac{\sd |\mathbf{q}|}{(2\pi)^3}e^{i|\mathbf{q}|r\cos\theta - \epsilon|\mathbf{q}|r}|\mathbf{q}|^{n+2}\sin\theta,
\end{equation}
Integrating this with $\epsilon\longrightarrow 0$, we can define an identity valid for \emph{odd} integers $n$, satisfying $n \geq -1$
\begin{equation}\label{key}
\int \frac{\sd^3\mathbf{q}}{(2\pi)^3}e^{i\mathbf{q}\cdot \mathbf{r}}|\mathbf{q}|^n =
\frac{(n+1)!}{2\pi^2r^{3+n}}\sin\left(\frac{3\pi n}{2}\right).
%\frac{2\Gamma[2+n]\sin\left(\frac{3n\pi}{2}\right)}{r^{3+n}},
\end{equation}
Including a log piece and repeating the same procedure yields a similar but unfortunately more unwieldy identity, and so we simply note only the following identities
\begin{subequations}
\begin{align}\label{key}
\int \frac{\sd^3\mathbf{q}}{(2\pi)^3}e^{i\mathbf{q}\cdot \mathbf{r}}|\textbf{q}|^4\log(\textbf{q}^2) &= -\frac{60}{\pi r^7},\\[1em]
\int \frac{\sd^3\mathbf{q}}{(2\pi)^3}e^{i\mathbf{q}\cdot \mathbf{r}}|\textbf{q}|^6\log(\textbf{q}^2) &= \frac{2520}{\pi r^9}.%,\\[1em]
%\int \frac{\sd^3\mathbf{q}}{(2\pi)^3}e^{-i\mathbf{q}\cdot \mathbf{r}}|\textbf{q}|^3\log(\textbf{q}^2) &= \frac{50 - 24\gamma_E - \log(r^{24})}{\pi r^6},
\end{align}
\end{subequations}
%where $\gamma_E$ is the Euler–-Mascheroni constant.

%\begin{equation}\label{key}
%F[r,n] = \int \frac{d^3q}{(2\pi)^3}e^{-i\vec{q}\cdot \vec{r}}|\vec{q}|^n =
%\begin{cases}
%\frac{2(n+1)!\sin\left(\frac{3n\pi}{2}\right)}{r^{3+n}} & n \text{ odd, } n \geq -1 \\
%0 &\text{otherwise}.
%\end{cases}
%%\frac{2\Gamma[2+n]\sin\left(\frac{3n\pi}{2}\right)}{r^{3+n}},
%\end{equation}
%\nocite{*}
%\printbibliography
\bibliographystyle{JHEP}
\bibliography{bib2}
\end{document}